\documentclass[aps,twocolumn,preprintnumbers,nofootinbib,superscriptaddress]{revtex4-1}
\usepackage{times}
\usepackage{amsmath}
\usepackage{amssymb}
\usepackage{graphicx}
\usepackage{subfigure,accents}
\usepackage{booktabs}
\usepackage{rotating}
\usepackage{longtable}
\usepackage{bm}
\usepackage[pagebackref=false,colorlinks=,linkcolor=blue,citecolor=blue]{hyperref}
\usepackage{epstopdf}
 \pdfoutput=1
 \usepackage[utf8]{inputenc}
\usepackage{cases}
\usepackage{amsmath,float}
\usepackage{amssymb}
\usepackage{amsfonts}
\usepackage{amssymb}
\usepackage{dcolumn}
\usepackage{bm}
\usepackage{bbm}
\usepackage{graphicx}
\usepackage{xcolor}
\usepackage{array}
\usepackage{subfigure}
\usepackage{hyperref}
\usepackage{wasysym}

\newcommand{\be}{\begin{equation}}
\newcommand{\ee}{\end{equation}}
\newcommand{\ba}{\begin{eqnarray}}
\newcommand{\ea}{\end{eqnarray}}

\newcommand{\gsim}{\mathrel{\hbox{\rlap{\lower.55ex \hbox {$\sim$}}
                   \kern-.3em \raise.4ex \hbox{$>$}}}}
\newcommand{\lsim}{\mathrel{\hbox{\rlap{\lower.55ex \hbox {$\sim$}}
                   \kern-.3em \raise.4ex \hbox{$<$}}}}

\hypersetup{colorlinks=true,
            breaklinks=true,
            pdfstartview=Fit,
            linkcolor=blue,
            citecolor=blue,
            urlcolor=blue}

 \usepackage{graphicx}

\usepackage{amsmath}

\usepackage{braket}

\newcommand{\bw}{\begin{widetext}}
\newcommand{\ew}{\end{widetext}}

\newcommand{\diag}{\ensuremath{ \text{diag} }}

\def\ber{\begin{eqnarray}}
\def\eer{\end{eqnarray}}
\def\beq{\begin{equation}}
\def\eeq{\end{equation}}

\begin{document}
\title{Entropic corrections to Friedmann equations and bouncing universe due to the zero-point length}
\author{Kimet Jusufi}
\email{kimet.jusufi@unite.edu.mk}
\affiliation{Physics Department, State University of Tetovo,
Ilinden Street nn, 1200, Tetovo, North Macedonia}
\author{Ahmad Sheykhi}
\email{asheykhi@shirazu.ac.ir}
\affiliation{Physics
Department and Biruni Observatory, College of Sciences,\\
Shiraz University, Shiraz 71454, Iran}

\begin{abstract}
We employ Verlinde's entropic force scenario to extract the
modified Friedmann equations by taking into account the zero-point
length correction to the gravitational potential. Starting from
the modified gravitational potential due to the zero-point length,
we first find the logarithmic corrections to the entropy
expression and then we derive the modified Friedman equations.
Interestingly enough, we observe that the corrected Friedmann
equations are similar to the Friedmann equations in braneworld
scenario. In addition, from the corrected Friedmann equations, we
pointed out a possible connection to the GUP principle which might
have implications on the Hubble tension. To this end, we discuss
the evolution of the scale factor under the effect of zero-point
length. Finally, having in mind that the minimal length is of the
Planck order, we obtain the critical density and the bouncing
behavior of the universe with a critical density and a minimal
scale factor of the order of Planck length.
\end{abstract}
\maketitle

\section{Introduction}
Since the discovery of black holes thermodynamics in $1970$'s
\cite{Haw}, physicists have been speculating that there should be
a profound connection between the gravitational field equations
and the law of thermodynamics. This is due to the fact that
thermodynamic quantities such as entropy and temperature are,
respectively, proportional to the horizon area and surface
gravity, which are pure geometrical quantities. Jacobson was the
first who disclosed that Einstein field equation of gravity is
indeed an equation of state for the spacetime \cite{Jac}.
According to Jacobson's argument, one can derive the hyperbolic
second order partial differential equations of general relativity
by starting from the Clausius relation $\delta Q=T\delta S$,
together with the relation between horizon area and entropy
\cite{Jac}. Jacobson's discovery was a great step toward
understanding the nature of gravity and supports the idea that
gravitational field equations are nothing but the first law of
thermodynamics for the spacetime. The profound connection between
the first law of thermodynamics and the gravitational field
equations were also generalized to other gravity theories
including $f(R)$ gravity \cite{Elin}, Gauss-Bonnet gravity, the
scalar-tensor gravity, and more general Lovelock gravity
\cite{Cai1,Pad}. In the cosmological background, it has been shown
that the differential form of the Friedmann equation on the
apparent horizon can be rewritten in the form of the first law of
thermodynamics and vice versa
\cite{Cai2,Cai3,CaiKim,Wang,Cai33,Shey0,Shey1,Shey2,Shey3}.
Although Jacobson's derivation is logically clear and
theoretically sound, the statistical mechanical origin of the
thermodynamic nature of gravity remains obscure.

The next great step towards understanding the nature of gravity
put forwarded by Verlinde who claimed that gravity is not a
fundamental force and can be regarded as an entropic force
\cite{Ver}. Verlinde proposal based on two principles. The
equipartition law of energy for the degrees of freedom of the
system and the holographic principle. Using these two principles,
he derived the Newton's law of gravity, the Poisson equations and
in the relativistic regime, the Einstein field equations of
general relativity. Similar discoveries were also made by
Padmanabhan \cite{TPad3} who observed that the equipartition law
for horizon degrees of freedom combined with the Smarr formula
leads to the Newton's law of gravity. This may imply that the
entropy is to link general relativity with the statistical
description of unknown spacetime microscopic structure when the
horizon is present.

It is important to note that Verlinde's proposal changed our
understanding on the origin and nature of gravity, but it
considers the gravitational field equations as the equations of an
emergent phenomenon and leave the spacetime as a background geometry which already exists. In line with studies to
understand the nature of gravity, Padmanabhan \cite{PadEm} argued
that the spacial expansion of our Universe can be understood as
a consequence of the emergence of space. Equating the difference
between the number of degrees of freedom in the bulk and on the
boundary with the volume change, he extracted the Friedmann
equation describing the evolution of the Universe \cite{PadEm}.
The idea of emergence spacetime was also extended to Gauss-Bonnet,
Lovelock gravities \cite{CaiEm,Yang,Sheyem}.

In the present work, we adopt the viewpoint of Verlinde and
consider gravity as an entropic force caused by the changes in the
information of the system. Using this scenario, we shall
investigate the effects of zero-point length corrections to the
gravitational potential on the cosmological field equations. The
concept of duality and zero point length was derived on the
framework of quantum gravity by Padmanabhan \cite{padma}. In his
reasoning, the spacetime manifold can be taken as a large distance
limit of the main quantum spacetime. In this manner, the discrete
to continuum transition should have a memory of the fluctuations
of the quantum spacetime. It was recently used to obtain black
hole solutions \cite{td1,td2,Nicolini:2022rlz,td3}. In order to incorporate such
quantum gravity effects we need to know the modified entropy.
Toward this goal, we shall use Verlinde's entropic force scenario
\cite{Ver} to obtain the corrected entropy.

This paper is outlined as follows. In Section II, we derive the
corrected entropy and the modified Friedmann equations. In Section
III, we discuss the evolution of the scale factor under the effect
of zero-point length. In Section IV, we explore the bouncing
behavior of the modified Friedmann equations. We comment on our
results in Section IV. Through the paper we  set $G=c=\hbar=1$.
\section{Entropic Corrections to Friedmann Equations}
According to Verlinde's argument, when a test particle or
excitation moves apart from the holographic screen, the magnitude
of the entropic force on this body has the form
\begin{equation}
F\triangle x=T \triangle S,
\end{equation}
in this equation $\triangle x$ gives the displacement of the particle from the
holographic screen, on the other hand $T$ and $\triangle S$ are the
temperature and the entropy change on the screen, respectively. Verlinde's derivation of Newton's law of gravitation at the very
least offers a strong analogy with a well understood statistical
mechanism. It is important to note that in Verlinde discussion,
the black hole entropy $S$ plays a crucial role in the derivation
of Newton's law of gravitation. Indeed, the derivation of Newton's
law of gravity depends on the entropy-area relationship $S=A/4$ of
black holes in Einstein's gravity, where $A =4\pi R^2$ represents
the area of the horizon. However, this definition can be modified
from the inclusion of quantum effects, here we shall use the
following modification \cite{Sheykhi:2010wm,Sheykhi:2021fwh}
\begin{eqnarray}
 S=\frac{A}{4}+\mathcal{S}(A).
\end{eqnarray}

It can be seen that if we take the second term zero, the standard
Bekenstein-Hawking result is reproduced. Assuming a test mass $m$
with a distance $\triangle x = \eta \lambda_m$ away from the
surface $\Sigma$, where $\lambda_{m}$ is the reduced Compton
wavelength of the particle given by $\lambda_{m}=1/m$ in natural
units, $\eta$ is some constant of proportionality. then the
entropy of the surface changes by one fundamental unit $\triangle
S$ fixed by the discrete spectrum of the area of the surface via
the relation
\begin{equation}
\label{S3}
 d S=\frac{\partial S}{\partial A}d A=\left[\frac{1}{4}+\frac{\partial \mathcal{S}}{\partial A}\right]d
 A.
\end{equation}
Here we note that the energy of the surface $\mathcal {S}$ is identified with the
relativistic rest mass $M$ of the source mass:
\begin{equation}
\label{Ec} E=M.
\end{equation}
On the surface $\Sigma$, we can relate the area of the surface to
the number of bytes according to
\begin{equation}
\label{AQN}
 A=QN,
 \end{equation}
where $Q$ is a fundamental
constant and $N$ is the number of bytes. Let us assume tha the temperature
on the surface is $T$, by means of the equipartition law
of energy \cite{Pad1}, we get the total energy on the surface via
\begin{equation}
\label{E}
 E=\frac{1}{2}Nk_B T.
 \end{equation}
We also need the force, which, according to this picture, it is the entropic force obtained from the thermodynamic
equation of state
\begin{equation}\label{F2}
F=T \frac{\triangle S}{\triangle x},
\end{equation}
where $\triangle S$ is one fundamental unit of entropy when
$|\triangle x|= \eta \lambda_m$, and the entropy gradient points
radially from the outside of the surface to inside. Note that we
have $\triangle N=1$, hence from (\ref{AQN}) we have $\triangle
A=Q$. Now, we are in a position to derive the entropic-corrected
Newton's law of gravity. Combining Eqs. (\ref{S3})- (\ref{F2}), we
can get $F$ using $\Delta S \simeq dS$, furthermore as we noted
$\Delta x=\eta \lambda_m=\eta/m $  and $T$ is found from Eq. (6),
where $N=A/Q$ and $T=MQ/(2\pi k_B R^2)$, we get
\begin{equation}\label{F3}
F=-\frac{Mm}{R^2}\left(\frac{Q^2}{2\pi k_B
\eta}\right)\left[\frac{1}{4}+\frac{\partial \mathcal{S}}{\partial
A}\right]_{A=4\pi R^2},
\end{equation}
This is nothing but the Newton's law of gravitation to the first
order provided we define $\eta=1/8\pi k_B$, we get $Q^2=1$. Thus
we reach
\begin{equation}\label{F4}
F=-\frac{Mm}{R^2}\left[1+4\,\frac{\partial \mathcal{S}}{\partial A}\right]_{A=4\pi R^2},
\end{equation}
In T-duality, it was argued that the gravitational potential due to the zero-point length is modified as \cite{td1,td2}
\begin{eqnarray}
\phi(r)=-\frac{M}{\sqrt{r^2+l_0^2}}|_{r=R},
\end{eqnarray}
where $l_0$ is the zero-point length and its value is expected to
be of Planck length (see, \cite{td1}). Now by using the relation
$\vec{F}=-m\nabla \phi(r)|_{r=R}$, we obtain the modified Newton's
law at of gravitation as
\begin{equation}\label{F5}
F=-\frac{Mm}{R^2}\left[1+\frac{l_0^2}{R^2}\right]^{-3/2}.
\end{equation}
Thus, with the  correction
in the entropy expression, we see that
\begin{eqnarray}
1+\left(\frac{1}{2 \pi
R}\right)\frac{d\mathcal{S}}{dR}=\left[1+\frac{l_0^2}{R^2}\right]^{-3/2}
\end{eqnarray}
Solving the above equation, for the entropy we obtain
\begin{eqnarray}
S&=&\pi R^2+\mathcal{S}=\pi R^2
\left(1+\frac{l_0^2}{R^2}\right)^{-1/2}+3\pi l_0^2
\left(1+\frac{l_0^2}{R^2}\right)^{-1/2}\nonumber\\&&-3 \pi l_0^2
\ln(R+\sqrt{R^2+l_0^2}),
\end{eqnarray}
It is very interesting to see that we obtained the log corrections
to the entropy in accordance with quantum effects. This is the
first important results in the present work. Let us now extend our
discussion to the cosmological setup. Assuming the background
spacetime to be spatially homogeneous and isotropic which is given
by the Friedmann-Robertson-Walker (FRW) metric
\begin{equation}
ds^2={h}_{\mu \nu}dx^{\mu} dx^{\nu}+R^2(d\theta^2+\sin^2\theta
d\phi^2),
\end{equation}
where $R=a(t)r$, $x^0=t, x^1=r$, the two dimensional metric
\begin{equation}
    h_{\mu \nu}=\diag(-1, a^2/(1-kr^2))
\end{equation}
Here $k$ denotes the
curvature of space with $k = 0, 1, -1$ corresponding to flat, closed, and open universes, respectively. The dynamical apparent
horizon, a marginally trapped surface with vanishing expansion, is
determined by the relation
\begin{eqnarray}
h^{\mu
\nu}\partial_{\mu}R\partial_{\nu}R=0,
\end{eqnarray}
A simple calculation gives
the apparent horizon radius for the FRW universe
\begin{equation}
\label{radius}
 R=ar=\frac{1}{\sqrt{H^2+k/a^2}}.
\end{equation}
For the matter source in the FRW universe we shall assume a perfect
fluid described by the stress-energy tensor
\begin{equation}\label{T}
T_{\mu\nu}=(\rho+p)u_{\mu}u_{\nu}+pg_{\mu\nu}.
\end{equation}
On the other hand, the total mass $M = \rho V$ in the region
enclosed by the boundary $\mathcal S$ is no longer conserved, one can compute the change in the total mass using the pressure
$dM = -pdV$, and this leads to the continuity equation
\begin{equation}\label{Cont}
\dot{\rho}+3H(\rho+p)=0,
\end{equation}
with $H=\dot{a}/a$ being the Hubble parameter. Let us now derive the dynamical equation for Newtonian cosmology. Toward this goal, let us consider a compact spatial region $V$ with a compact boundary
$\mathcal S$, which is a sphere having radius $R= a(t)r$, where $r$ is a dimensionless quantity. Going back and combining the second law of Newton for the test
particle $m$ near the surface, with gravitational force (\ref{F5})
we obtain
\begin{equation}\label{F6}
F=m\ddot{R}=m\ddot{a}r=-\frac{Mm}{R^2}\left[1+\frac{l_0^2}{R^2}\right]^{-3/2},
\end{equation}
We also assume $\rho=M/V$ is the energy density of the matter
inside the the volume $V=\frac{4}{3} \pi a^3 r^3$. Thus, Eq.
(\ref{F6}) can be rewritten as
\begin{equation}\label{F7}
\frac{\ddot{a}}{a}=-\frac{4\pi
}{3}\rho \left[1+\frac{l_0^2}{R^2}\right]^{-3/2},
\end{equation}
this result represent the entropy-corrected dynamical equation for
Newtonian cosmology. In order to derive the Friedmann equations of FRW universe in general relativity, we can use the active gravitational mass $\mathcal M$, rather than the total mass $M$. It follows that, due to the
entropic corrections terms via the zero-point length, the active gravitational mass
$\mathcal M$ will be modified. Using Eq.
(\ref{F7}) and replacing $M$ with $\mathcal M$, it follows
\begin{equation}\label{M1}
\mathcal M =-\ddot{a}
a^2r^3\left[1+\frac{l_0^2}{R^2}\right]^{3/2}
\end{equation}
In addition, for the active gravitational mass we can use the definition
\begin{equation}\label{Int}
\mathcal M =2
\int_V{dV\left(T_{\mu\nu}-\frac{1}{2}Tg_{\mu\nu}\right)u^{\mu}u^{\nu}}.
\end{equation}
From here it is not difficult to show the following result
\begin{equation}\label{M2}
\mathcal M =(\rho+3p)\frac{4\pi}{3}a^3 r^3.
\end{equation}
By means of Eqs. (\ref{M1}) and (\ref{M2}) we  find
\begin{equation}\label{addot}
\frac{\ddot{a}}{a} =-\frac{4\pi
}{3}(\rho+3p)\left[1+\frac{l_0^2}{R^2}\right]^{-3/2}.
\end{equation}
This is the modified acceleration equation for the dynamical
evolution of the  FRW universe. To simplify the work, since $l_0$ is a very small number, we can consider a series expansion around $l_0$ via
\begin{eqnarray}
\left[1+\frac{l_0^2}{r^2 a^2}\right]^{-3/2}=1-\frac{3}{2}\frac{l_0^2}{r^2 a^2}+...
\end{eqnarray}
For the Friedmann equation we therefore obtain
\begin{equation}\label{addot0}
\frac{\ddot{a}}{a}=- \left(\frac{4 \pi  }{3}\right)(\rho+3p)\left[1-\frac{3}{2} \frac{l_0^2}{r^2 a^2}+... \right] .
\end{equation}
Next, by multiplying $2\dot{a}a$ on both sides of Eq. (\ref{addot}), and by means of continuity equation (\ref{Cont}), we obtain
\begin{equation}\label{Fried1}
\dot{a}^2+k =\frac{8\pi
}{3}\int d(\rho a^2)\left[1-\frac{3}{2} \frac{l_0^2}{r^2 a^2}+... \right],
\end{equation}
where $k$ is a constant of integration and physically characterizes the curvature of space.
Using the expression for density
\begin{eqnarray}
\rho=\rho_0 a^{-3 (1+w)},
\end{eqnarray}
we obtain in leading order terms
\begin{equation}\label{Fried1}
H^2+\frac{k}{a^2} =\frac{8\pi
}{3}\rho\left[1-\frac{l_0^2}{2}\left(H^2+\frac{k}{a^2}\right)\frac{1+3 \omega}{1+\omega}+...\right].
\end{equation}
It is convenient, to simplify the above equation by rewriting it
as
\begin{equation}\label{Fried2}
H^2+\frac{k}{a^2} =\frac{8\pi }{3}\rho\left[1-\Gamma \rho\right],
\end{equation}
where $\Gamma$ is a constant defined as
\begin{equation}\label{Gamma}
\Gamma\equiv\frac{4l_0^2\pi }{3}\left(\frac{1+3
\omega}{1+\omega}\right).
\end{equation}
We see that in the limit $l_0 \to 0$ the standard Friedmann
equation is obtained. Eq. (\ref{Fried2}) is similar to the
Friedmann equations in braneworld scenario \cite{bw1} (see, in
particular \cite{Shey1,bw2}).

This is one of the most interesting results found in the present paper, and
the correspondence or the link between these models can be seen from the fact that string theory
has a T-duality symmetry relating circle compactifications of large and small radius.  Put in other words, T-duality, identifies string theories on higher-dimensional spacetimes with mutually inverse compactification radii. The idea of extra spatial dimensions is not new and goes back to Kaluza and Klein. As we noted, these extra dimensions can be compactified for example on a small enough radius (in the traditional Kaluza-Klein sense), however, these extra dimensions can be also large, in the sense that the ordinary matter is confined onto a three-dimensional subspace, known as the
brane, embedded in a larger space, known as the bulk. The idea of these large extra dimension, was used in braneworld cosmology to explain the weakness of gravity relative to the other fundamental forces of nature.

From the modified equations we can see that there
is an apparent singularity encoded in $\Gamma$ and an asymptotic behavior is obtained when $\omega
\to -1$ then $\Gamma\to \infty$, which signals a phase transition of the universe.To summarize, we derived the entropy-corrected Friedmann equation of
FRW universe by considering gravity as an entropic force caused by
changes in the information associated with the positions of
material bodies. From the entropy-corrected
Friedmann equation of the FRW universe we expect these corrections to play an important role in the early stage of the universe. Furthermore, we can derive the modified
Raychaudhuri equation using
\begin{eqnarray}
\dot{H}=-H^2+\frac{\ddot{a}}{a}.
\end{eqnarray}
This equation can also be written as
\begin{equation}
\dot{H}=\frac{k}{a^2}-\frac{8\pi }{3}\Big\{\rho\left[1-\Gamma \rho\right]
+\frac{\rho+3p}{2 \left[1+l_0^2 \left(H^2+\frac{k}{a^2}\right)\right]^{3/2}}\Big\}.
\end{equation}
In addition, there is also the deceleration parameter which
is defined as
\begin{equation}
q=-1-\frac{\dot{H}}{H^2},
\end{equation}
where depending on the sign of $q$ we can have deceleration or
acceleration scenario, respectively. Let us show another interesting results from Eq. (\ref{Fried2}) and for simplicity we choose a flat universe with $k=0$ along with the definition $H_0^2=8 \pi \rho/3$. From Eq. (\ref{Fried2}) we obtain
\begin{eqnarray}
H=H_0\left(1-\alpha H_0^2 l_0^2\right)^{1/2},
\end{eqnarray}
where $\alpha$ is some constant of proportionality. Considering a series expansion around $l_0$, we obtain
\begin{eqnarray}
H \simeq H_0 \left(1-\Delta_{GUP} H_0^2\right),
\end{eqnarray}
where $H$ has the form of a Generalized Uncertainty Principle (GUP) modified Hubble parameter with $\Delta_{GUP}=\alpha l_0^2/2$. It is expected that the Cosmic Microwave Background  (CMB) carries fingerprints of quantum gravity therefore for $H$ we can take the value reported by the Planck collaboration, $H=67.40 \pm 0.50$ km s$^{-1}$ Mpc$^{-1}$ \cite{Planck:2018vyg}.  On the other hand $H_0$ can be viewed as the unmodified Hubble parameter, and we can take the value reported by the Hubble Space Telescope (HST), $H_0=74.03 \pm 1.42$ km s$^{-1}$ Mpc$^{-1}$ \cite{Riess:2019cxk}. This problem of disagreement between the current value of Hubble parameter is known as the Hubble tension in modern cosmology. The last equation can also be written as
\begin{eqnarray}\label{GUP}
\frac{H_0-H}{H_0^3}=\Delta_{GUP}.
\end{eqnarray}
Quite interesting, if we further replace $\Delta_{GUP}=432 \lambda_2 a^4$, the last equation coincides exactly with the GUP modified Hubble parameter given in  \cite{Aghababaei:2021gxe}. The GUP modified Hubble parameter can have interesting implications in cosmology. In particular, according to this picture, one way to see the Hubble tension is to associate the corrected CMB Hubble parameter $(H)$ to be the one measured by the Planck collaboration which uses the CMB data. This is related to the fact that $H$ is expected to encode the quantum gravity effects measured by CMB. On the other hand, we associate the unmodified Hubble parameter $(H_0)$, to be measured by the HST group which uses the SNeIa data. That being said, one can now use the values for the Hubble parameter given in  \cite{Planck:2018vyg,Riess:2019cxk} along with Eq. \eqref{GUP}, to constrain the GUP parameter  $\Delta_{GUP}$ (see, \cite{Aghababaei:2021gxe}).  
Here we note that other authors as well studied the cosmological Hubble tension using the Heisenberg uncertainty, for example, the interested reader can see \cite{Capozziello:2020nyq}.
\section{Evolution of the scale factor}
Let us define the energy density due to the curvature via
\begin{equation}
\rho_{\rm{curv}}=-\frac{3 k}{8 \pi  a^2},
\end{equation}
then, we can rewrite the Friedmann equation as
\begin{equation}
H^2=\frac{8\pi  }{3}\left[ \rho (1-\Gamma
\rho)+\rho_{\rm{curv}}\right].
\end{equation}
We are going to consider first the case of flat universe
($\rho_{\rm{curv}}=0$). To solve the above equation we can take
$\rho=\rho_0 a^{-n}$. For a matter dominated universe we have
$n=3$, which leads to
\begin{equation}
a(t) \sim \left( t^2+\frac{\Gamma}{6 \pi  }\right)^{1/3}.
\end{equation}
In the limit $\Gamma \to 0$ we get the standard law $a(t) \sim  t ^{2/3}$.
For a radiation dominated we have $n=4$, which leads to
\begin{equation}
a(t) \sim \left( t^2+\frac{3 \Gamma}{32 \pi }\right)^{1/4}.
\end{equation}
In the limit $\Gamma \to 0$, again we get the standard law $a(t) \sim  t ^{1/2}$.
Finally, we can find the scale factor for a vacuum dominated universe if we can set $n=0$, which leads to
\begin{equation}
a(t) \sim e^{C \sqrt{(1-\Gamma \rho_0)} t},
\end{equation}
where $C$ is a constant. In the limit $\Gamma \to 0$, we can
identify $C=H$, yielding $a(t) \sim e^{H t}$. There is an
interesting special case if we consider a universe with curvature
dominated energy and say we take $\rho=3k/8 \pi  a^2$, which
cancels out the term $\rho_{\text{curv.}}$, then we get
\begin{equation}
a(t) \sim \left(-k^2 \Gamma t^2\right)^{1/4},
\end{equation}
which make sense only for $k=\pm 1$, along with a specific domain for the parameter $\omega$. Note also that such a scale
factor, is absent in the classical picture with the limit $l_0 \to 0$.
\section{Bouncing behavior of the modified Friedmann equations}
In this section, we are going to use some of these equations, to
study the bouncing behavior of the modified Friedmann equations.
According to the bouncing paradigms, there might be a nonsingular
connection between the contraction phase and the expansion phase.
During this phase change the universe goes through its minimal
value with non-vanishing value referred as a critical point.
During this process, quantum effects can play a dominant role and
prevent the universe from collapsing into a singularity and then
drive our universe to accelerate expansion. Mathematically, at the
critical point we can write the minimal nonzero value ($a_0>0$)
which has $H|_{a=a_0}=0$ along with the condition $\ddot{a}_0>0$
(see, \cite{Ling:2009wj,Pan:2015tza}). Setting $H|_{a=a_0}=0$,
from Eq. \eqref{Fried2} we find two branches of solution for the
critical density
\begin{equation}
\rho_{\text{crit.}}=\frac{1}{2 \Gamma}\left(1\pm \sqrt{1-\frac{3 \Gamma k}{2 \pi a_0^2}}\right).
\end{equation}
Here we further impose the condition $1-3 \Gamma k/2  \pi a_0^2
\geq 0$, from where we find the minimal quantity for the scale
factor
\begin{equation}
a_0=\pm \sqrt{\frac{3 \Gamma k}{2 \pi}}.
\end{equation}
Furthermore, we can take as a physical solution only the positive
branch and having in mind Eq. \eqref{Gamma} we obtain
\begin{equation}
a_0= \sqrt{2}\,l_0 \sqrt{\frac{ k (1+3 \omega)}{1+\omega}}.
\end{equation}
Since $l_0$ is of the order of Planck length, we see that the
universe has bouncing behavior at Planck length. As a first
example, if we set $k=0$, we see that the first condition $a_0>0$
is not satisfied, hence the bouncing behavior for the flat
universe is absent. For a closed universe with $k=+1$, we get
\begin{equation}
a_0= \sqrt{2}\,l_0 \sqrt{\frac{1+3 \omega}{1+\omega}},
\end{equation}
provided $(1+3 \omega)/(1+\omega)>0$. On the other hand for $k=-1$, in order to get $a_0>0$, we obtain
\begin{equation}
a_0= \sqrt{2}\,l_0 \sqrt{-\frac{ (1+3 \omega)}{1+\omega}},
\end{equation}
provided $(1+3 \omega)/(1+\omega)<0$. Now if we further take at the minimal value $a=a_0$, from Eq. \eqref{addot0} we obtain
\begin{equation}
\frac{\ddot{a}_0}{a_0}=- \left(\frac{4 \pi   }{3}\right)(\rho_{\text{crit.}}+3p_{\text{crit.}})
\left[  1-\frac{3}{4}\frac{1+\omega}{1+3 \omega}+... \right] .
\end{equation}
In order to get the condition $\ddot{a}_0>0$, we can see that
there are two possibilities: First possibility is to set
$\rho_{\text{crit.}}+3p_{\text{crit.}}<0$ and $1-3
(1+\omega)/4(1+3\omega)>0$. Second possibility, however, is to
take $\rho_{\text{crit.}}+3p_{\text{crit.}}>0$ and $1-3
(1+\omega)/4(1+3\omega)<0$. In what follows we shall use the
second possibility. Let us take now for example the closed
universe with $k=+1$, then from the condition $a_0>0$, it follows
the interval $\omega \in (-\infty, -1) \cup (-1/3, \infty)$, while
from the condition $\ddot{a}_0>0$, we get the interval $\omega \in
(-1/3, -1/9)$.   This means that only in the interval $\omega \in
(-1/3, -1/9)$, a closed universe, has bouncing behavior. Finally,
let us consider the scenario having a universe with $k=-1$, then
from the  condition $a_0>0$ we get the interval $\omega \in (-1,
-1/3)$, while from the condition $\ddot{a}_0>0$ it follows $\omega
\in (-1/3, -1/9)$. We conclude that, the bouncing behavior not
possible in this case. We found that there is a great level of
similarity between the results found in the present work and the
bouncing behavior due to the modified dispersion relation
\cite{Ling:2009wj,Pan:2015tza}. In particular, it was found for
the bouncing condition $\ddot{a}_0/a_0 =1/(\eta_0^2l_p^2)$, where
$\eta_0$ is some dimensionless parameter which is bigger than zero
and $l_p$ is the Planck length.  This suggest that a closed
universe may perform bouncing in the high energy limit when the
role of zero-point length is expected to be important.
We can show that this conclusion follows directly from our Eq. (50), where we have the proportionality $\ddot{a}_0/a_0 \sim \rho_{\text{crit.}}+3 p_{\text{crit.}}$. Furthermore, in this regime, we have
\begin{equation}
\frac{\ddot{a}_0}{a_0} \sim \frac{1}{\Gamma} \sim \frac{1}{l_0^2},
\end{equation}
where the zero-point length can be identified with the Planck length $l_0 \sim l_p$ (see for more details \cite{td1}). Solving the last equation using $a_0(t) \sim e^{\lambda t}$, we obtain for the scale factor
\begin{equation}
a_0(t) \sim A_1\exp \left(\frac{\mathcal{C}}{l_0}t  \right)+A_2\exp \left(-\frac{\mathcal{C}}{l_0}t  \right),
\end{equation}
where $\mathcal{C}$, $A_1$ and $A_2$ are some constants. Having in mind that there is a minimal value for $a_0$ (see Eq. (47)), at the initial time $t=0$, we have to set $A_2=0$, then we end up with
\begin{equation}
a_0(t) \sim A_1\exp \left(\frac{\mathcal{C}}{l_0}t  \right),
\end{equation}
where $t$ is the time of universe bounce and $A_1$ is proportional to $l_0$. By taking $\mathcal{C}\sim l_0^{-1}$ and $l_0 \sim l_p \sim 10^{-35}$ m, one can obtain the huge expansion of the early universe via exponential function.  It is interesting that Eq. (52) may suggest that when the universe reaches the minimal scale length, such a state is not stable and, we end up with the bouncing universe. The full mechanism is yet to be found and it is outside the scope of this work. The second term in this equation can describe the inverse scenario, namely, a state of matter that undergoes a collapsing processes. Due to the quantum gravitational bounce, or zero point length effect, there is no singularity in such a universe  described by the critical density. Let us now show that indeed the curvature scalars are regular. Using the Einstein field equations, one can find the Ricci scalar
\begin{eqnarray}
 \mathcal{R}=8\pi \rho(r).
\end{eqnarray}

Considering now the condition $\rho(r)=\rho_{\text{crit.}}$, at the minimal value for $a_0$, it can be shown that
\begin{eqnarray}
\mathcal{R}=\frac{8\pi}{2\Gamma}=\frac{3}{l_0^2} \frac{1+\omega}{1+3 \omega}.
\end{eqnarray}

This shows that there is no singularity in the expression for the Ricci scalar, provided $l_0$ is not zero along with the condition $\omega \in (-\infty, -1) \cup (-1/3, \infty)$. Note here that as was shown in \cite{Das:2008kaa,Ali:2009zq}, one can define the minimal length of GUP via $l_{0}=\alpha_0 l_{p}$ [where $l_{p}$ is the Planck length] so $l_{0}$ is proportional to $l_{p}$ and the limit to get standard results is to use $\alpha_{0}$. In particular, by taking the dimensionless GUP parameter $\alpha_{0} \to 0$, we obtain the standard result. However, there is an apparent singularity due to the parameter $\omega \to -1$, which signals that the the universe undergoes a phase transition in this stage. One can calculate one more scalar invariant, known as the Kretschmann scalar and can be computed via \cite{Kruglov:2020axn}
\begin{equation}
\mathcal{K} =(8\pi)^2\left[ \frac{5}{3}\rho_{\text{crit.}}^2+2 \rho_{\text{crit.}}p_{\text{crit.}} +3 p^2_{\text{crit.}} \right].
\end{equation}

From this equation, one can then show the following result
\begin{equation}
\mathcal{K} =\frac{3}{l_0^4} \frac{(1+\omega)^2\left(5+6 \omega +9 \omega^2 \right)}{(1+3 \omega)^2}.
\end{equation}
Again, we can see from the last equation that there is no
singularity due to $l_0$, but there are apparent singularities due
to the parameter $\omega$. As we already pointed out, these
singularities are a result of the phase transition of the
universe.
\section{Conclusions} \label{Con}
Adopting the concept of the zero-point length correction to the
gravitational potential, and employing the entropic force scenario
proposed by Verlinde, we computed the corrections to the Newton's
law of gravity as well as Friedmann equations. By assuming the
effect of zero-point length on the gravitational potential, we
find several important results.\\

Firstly, by using the modified Newton law of gravity and the
Verlindes entropic force, we found logarithmic correction terms to
the entropy and then we found that Friedman equations are indeed
affected by the zero-point length.

Secondly, from the modified Friedmann equations we found a
correspondence with the Friedmann equations in braneworld
scenario. In addition to that, we observed that from the corrected
Friedmann equations, we can obtain a form of GUP modified Hubble
parameter similar to one reported in \cite{Aghababaei:2021gxe} and
we found an upper bound for the GUP parameter.  This might have
interesting implications, in particular the GUP modified Hubble
parameter might shed some light on the Hubble tension problem in
cosmology \cite{Aghababaei:2021gxe}. Specifically, the corrected
CMB Hubble parameter $(H)$ encodes quantum gravity effects and is
expected to be measured by the Planck collaboration which uses the
CMB data, while the unmodified Hubble parameter is identified to
$(H_0)$, and is expected to be measured by the HST group which
uses the SNeIa data \cite{Dainotti:2021pqg,Dainotti:2022bzg}.

Thirdly, since the effect of $l_0$ is expected to play important
role in the early universe, we found a bouncing behavior at the
minimal value $a=a_0$ which is of the order of $l_0$. We have
elaborated the bouncing condition and it is shown that it is
possible only for the closed universe with $k=+1$. The bouncing
scenario is linked to the high energy limit when the role of
zero-point length is expected to be important. i.e., when
$\ddot{a}_0/a_0 \sim 1/l_0^2$. Similar results were reported in
\cite{Ling:2009wj,Pan:2015tza} where the bouncing behavior was
obtained from the modified dispersion relation. Finally, for the
expansion of universe in this regime we found the exponential law
$a_0(t) \sim \exp \left(\mathcal{C}t/l_0  \right)$, and the
curvature scalars are finite.

\acknowledgments{We are grateful to the referee for very careful
and constructive comments which helped us improve our paper
significantly.}


\end{document}